\documentclass[11pt,a4paper]{article}

\usepackage[a4paper,margin=0.75in]{geometry}
\usepackage{authblk}
\usepackage[authoryear]{natbib}
\usepackage[hidelinks]{hyperref}
\usepackage{float}
\usepackage{subcaption}
\usepackage{amsmath,amsfonts,bm}
\usepackage{dsfont}
\usepackage{multirow}
\usepackage{threeparttable}
\usepackage{booktabs}
\usepackage{graphicx}
\usepackage{array}
\usepackage{tabularx}
\usepackage{placeins}
\usepackage{cleveref}

\newlength{\tblwidth}
\AtBeginDocument{\setlength{\tblwidth}{0.96\textwidth}}

\title{Beyond Marginal Risk: ESG and Multidimensional Fragility in Equity Markets}
\author[1]{Minxuan Hu}
\author[2]{Jiayu Yi}
\author[3]{Ziheng Chen}
\author[4]{Wenxi Sun}
\author[5]{Qishi Zhan}
\affil[1]{Cornell Ann S. Bowers College of Computing and Information Science, Cornell University, USA}
\affil[2]{School of Social Sciences, Nanyang Technological University, Singapore}
\affil[3]{Department of Mathematics, University of Texas at Austin, USA}
\affil[4]{Krieger School of Arts and Sciences, Johns Hopkins University, USA}
\affil[5]{Department of Mathematical and Statistical Sciences, Marquette University, USA}
\date{}

\hypersetup{
  pdftitle={Beyond Marginal Risk: ESG and Multidimensional Fragility in Equity Markets},
  pdfauthor={Minxuan Hu, Jiayu Yi, Ziheng Chen, Wenxi Sun, Qishi Zhan}
}

\begin{document}
\maketitle
\thispagestyle{empty}

\begin{abstract}
Do ESG characteristics provide financial resilience beyond conventional risk, and is that inference stable to the financial state and ESG score being measured? We study a point-in-time panel of nonfinancial S\&P 500 constituents from January 2019 through November 2025. Once ex ante return, volatility, and liquidity exposures are modeled explicitly, the broad downside-resilience pattern attenuates and none of the three primary marginal ESG $\times$ Down tests survives Holm adjustment under either conventional or wild-cluster-bootstrap inference. A post-primary exact decomposition then shows that the negative Return+RVOL coefficient is almost entirely allocated to observations that also breach the CHL trading-cost tail, while the mutually exclusive Return+RVOL-only component is essentially zero. Finally, a paired standardized comparison on an identical sample shows that MSCI's downstream Industry Adjusted Score and upstream Weighted Average Key Issue Score imply materially different fitted associations in adverse market states: at a 5\% market decline, the difference is about one probability percentage point per score standard deviation. Sector-time, leave-one-month-out, and observed-change timing diagnostics preserve most of this contrast, although the result remains exploratory and inference is sensitive to multiplicity and market-state support. The evidence therefore favors a narrower interpretation of ESG resilience: the apparent association weakens once conventional risk is modeled, while the remaining relation depends on both the adverse state and the MSCI score field used to represent ESG.
\end{abstract}

\section{Introduction}
\label{sec:introduction}

Whether environmental, social, and governance (ESG) characteristics provide financial resilience is a central question in sustainable finance. Evidence from major stress episodes suggests that stronger stakeholder relationships or environmental and social profiles can coincide with better corporate and equity-market performance \citep{lins2017social,albuquerque2020resiliency,ding2021corporate,yahya2023resilience}. Successful ESG engagement has also been linked to lower downside risk \citep{hoepner2024engagement}, while more recent regime-based evidence suggests that any buffering effect can vary across market states \citep{duong2026buffer}. Other work is more cautious and finds that ESG does not necessarily protect equity values once conventional firm characteristics are made explicit \citep{demers2021esg}. These mixed findings motivate a more demanding finance question than whether high-ESG firms simply look safer: \emph{how much of apparent ESG resilience is incremental to conventional financial risk, where does any residual association reside, and does the answer depend on how ESG itself is measured?}

Three empirical issues are central to that question. First, firms with higher ESG scores may enter weak markets with lower beta, lower pre-existing volatility, or better liquidity. Without modeling those exposures directly, an ESG interaction with market conditions can be difficult to distinguish from ordinary financial risk sensitivity. Second, bad financial states are multidimensional. Price losses can coincide with volatility spikes and deteriorating tradability, particularly when market-wide conditions are unstable \citep{brunnermeier2009market,chordia2000commonality,amihud2002illiquidity,greenwood2011fragility}. A firm experiencing all three conditions occupies a different state from one experiencing only a return loss or volatility spike. Third, the ESG signal itself is not invariant. Scope, measurement, weighting, peer adjustment, and updating can materially change the score observed by investors \citep{berg2022aggregate,christensen2022corporate,avramov2022sustainable,msci2024methodology,msci2022issuerfaq}.

We address these issues in a monthly point-in-time panel of nonfinancial S\&P 500 constituents from January 2019 through November 2025. The primary ESG variable is MSCI's Industry Adjusted Score (IAS) observed in the exact preceding monthly snapshot. The marginal outcomes are monthly loss, log realized volatility, and the log close-high-low (CHL) spread estimator of \citet{abdi2017spread}, with larger values indicating worse conditions. All principal models include firm and month fixed effects, six standard firm controls, and two-way clustering by firm and month. The fully adjusted marginal models add the corresponding ex ante beta, volatility, or liquidity exposure and allow that exposure to vary with positive and negative market movements. The joint model includes all three exposure blocks simultaneously. Our objective is not to estimate a causal effect of ESG on financial fragility, but to characterize its conditional association with adverse financial outcomes after accounting for conventional financial exposures.

Three findings emerge. First, the broad resilience pattern weakens under the stronger financial benchmark. In the same-sample Joint80 model, adding beta, prior volatility, and prior liquidity attenuates the IAS interaction with negative market returns by about one third, while none of the three formally primary marginal ESG $\times$ Down tests survives Holm adjustment under either conventional or wild-cluster-bootstrap inference. Second, an exact post-primary decomposition localizes the residual Joint80 association. Almost the entire negative coefficient on the inclusive Return+RVOL state is allocated to observations that also breach the CHL trading-cost tail, while the mutually exclusive Return+RVOL-only component is essentially zero. The direct component contrast remains imprecise, so the decomposition identifies where the association is concentrated without establishing an independent liquidity mechanism.

Third, the financial relation is measurement dependent. On the identical 29,572-observation Joint80 sample, standardized MSCI Industry Adjusted Score (IAS) and the upstream Weighted Average Key Issue Score imply materially different adverse-state associations. At a 5\% market decline, the IAS-minus-WAKIS complete-slope difference is about one probability percentage point per score standard deviation. Targeted sector-time, leave-one-month-out, and observed-change timing diagnostics preserve most of that contrast. The comparison remains post-primary exploratory, and its inferential strength depends on the multiplicity family and the limited support for severe market declines. We therefore interpret it as evidence that the MSCI score field used to represent ESG matters for the estimated financial relation, not as identification of a particular production mechanism or proof that one score is superior.

The paper makes three contributions. First, it reframes ESG resilience against an explicit conventional-risk benchmark by allowing standard ex ante financial exposures to vary with market conditions. Second, it uses an exact decomposition of overlapping adverse states to locate the residual IAS association rather than infer a mechanism from separate regressions. Third, it treats ESG measurement as an economic object rather than a routine robustness choice: two aggregate score fields at different stages of the same provider's scoring architecture can imply materially different financial relationships even when the sample, outcome, controls, fixed effects, and inference design are held constant. Together, these results support a narrower and more disciplined interpretation of ESG resilience in which the conventional-risk benchmark, the adverse financial state, and the score construction are all part of the empirical claim being estimated.
\section{Related Literature and Economic Motivation}
\label{sec:literature}

\subsection{ESG, crisis performance, and financial resilience}

A large literature studies whether ESG or related stakeholder capital is associated with financial performance and risk \citep{friede2015esg,gillan2021firms,pastor2021sustainable,pedersen2021responsible}. The most relevant strand for this paper asks whether sustainability characteristics matter disproportionately in adverse states \citep{broadstock2021role}. \citet{lins2017social} show that firms with stronger social capital performed better during the global financial crisis, and \citet{albuquerque2020resiliency} document stronger COVID-period performance among firms with stronger environmental and social profiles. \citet{ding2021corporate} likewise show that pre-pandemic stakeholder and financial characteristics are associated with corporate resilience during COVID-19. In \textit{Global Finance Journal}, \citet{yahya2023resilience} reports positive associations between environmental and social performance and revenue, profitability, and valuation during the pandemic among Nordic firms. Related evidence from geopolitical stress links stronger CSR performance to less deterioration in market quality, particularly among foreign firms listed in the United States \citep{clanceyshang2024csr}.

The evidence is not uniformly favorable to an ESG-insurance interpretation \citep{bae2021csrcrisis}. \citet{demers2021esg} find that ESG did not immunize stocks during COVID-19 once other firm characteristics, particularly intangible assets, are considered. \citet{hoepner2024engagement} show that successful ESG shareholder engagement is followed by lower downside risk, whereas \citet{duong2026buffer} finds lower idiosyncratic volatility and tail losses for high-ESG portfolios in several regimes but not in the crash regime. ESG can also affect how investors respond to information rather than generate a simple unconditional return premium \citep{leite2023investors}. These differences are plausible because ESG can covary with stakeholder trust, information quality, operating flexibility, risk management, investor clientele, and conventional financial exposures. The empirical question is therefore not only whether ESG is associated with resilience, but whether it contributes information beyond characteristics that already make firms financially less exposed.

Prior work also tends to examine separate outcomes. ESG has been linked to stock-price crash risk \citep{kim2014csrcrash,bae2021esg,feng2022esgcrash}, liquidity \citep{luo2022esg}, and other risk measures. \citet{wang2023fragility} studies ESG and the Greenwood--Thesmar concept of stock-price fragility, which concerns sensitivity to nonfundamental investor trading demand. Our use of ``fragility'' is different: it denotes the simultaneous realization of adverse return, volatility, and trading-cost states within the same firm-month. We retain the marginal outcomes as the primary benchmark and use the joint state to ask where any residual association is concentrated.

\subsection{Multidimensional fragility and conventional risk exposure}

Finance theory and evidence provide strong reasons to treat bad states as multidimensional. Market and funding liquidity interact during stress \citep{brunnermeier2009market}, liquidity conditions comove across securities \citep{chordia2000commonality}, and illiquidity and liquidity risk matter for asset prices \citep{amihud2002illiquidity,acharya2005liquidity}. Fire-sale pressure can amplify price declines when investors transact in weak markets \citep{coval2007asset,falato2021firesale}, while concentrated ownership and common exposures can make individual securities vulnerable to correlated trading pressure \citep{greenwood2011fragility}. A month with a return loss and elevated volatility but stable trading costs is therefore not necessarily the same economic state as a month in which all three conditions deteriorate together.

Firm-level crisis responses can also differ across return and volatility dimensions \citep{han2026decomposingfirmlevelcrisisresponses}. This logic is related to work on systemic, common-tail, and multivariate risk \citep{kelly2014tail}. \citet{acharya2017measuring} emphasize that financial risk depends on how losses interact with bad aggregate states. More recently, \citet{bissoondoyal2026multivariate} construct a multivariate crash-risk measure and show that conditional crash exposure contains pricing information across international markets. We do not import a multivariate crash model into the ESG setting. Instead, transparent simultaneous-tail indicators and an exact decomposition of overlapping states allow us to distinguish a broad return-volatility condition from the more restrictive subset in which trading costs also deteriorate.

Conventional risk adjustment is central to that distinction. Downside-risk research shows that sensitivities in bad market states can differ from ordinary beta \citep{ang2006downside,levi2020betas}. 
Recent evidence from Chinese equity markets also identifies turnover and momentum as informative predictors of stock returns~\citep{han2026interpretablefactordecompositiondecision}. At the same time, a state-dependent ESG coefficient can proxy for lower beta, lower pre-existing volatility, or better liquidity. Our specifications therefore include those exposures directly and allow each to vary with negative and positive market movements. This sets a deliberately demanding benchmark for any incremental ESG association.

\subsection{ESG measurement architecture}

A separate literature questions whether an ESG score represents a single stable construct \citep{gibsonbrandon2021disagreement,billio2021inside,anselmi2025ratings}. \citet{berg2022aggregate} document substantial divergence across providers and attribute it to differences in scope, measurement, and weighting. \citet{christensen2022corporate} similarly show that corporate ESG evaluations depend on underlying measurement choices, while \citet{avramov2022sustainable} demonstrates that rating uncertainty can materially affect sustainable investment decisions. Recent evidence in \textit{Global Finance Journal} links ESG rating disagreement to firms' cost of equity capital \citep{lou2025disagreement}. Measurement instability may also arise within a provider when updating and weighting practices alter previously reported scores and the resulting empirical relationships \citep{sahin2023pitfalls}.

Our measurement analysis is narrower than a cross-provider comparison. We examine two aggregate score fields at different stages of the same MSCI scoring architecture: the downstream Industry Adjusted Score used in the primary analysis and the upstream Weighted Average Key Issue Score. The paired design holds the provider, firm-month sample, financial outcome, controls, fixed effects, and estimation structure constant and places both scores on comparable standardized scales. In the empirical sections, \emph{score architecture} refers to the combined construction and updating process represented by these fields. The resulting comparison asks whether a financial relation attributed to ``ESG'' is invariant even within one provider.

These literatures motivate a sequential empirical test. We first test whether apparent ESG resilience survives explicit conventional risk adjustment, then examine whether any remaining coefficient pattern is concentrated within overlapping adverse states, and finally assess whether that pattern depends on the MSCI score field used to represent ESG.
\section{Data and Variable Construction}
\label{sec:data}

\subsection{Sample and information timing}

We construct a monthly point-in-time panel of nonfinancial firms that were constituents of the S\&P 500 between January 2019 and November 2025. Historical constituent records \citep{wikipedia2026sp500} form the universe to which market, ESG, and accounting observations are linked. The primary window begins in 2019, when the linked sources provide stable joint coverage under the point-in-time construction. Financial firms are excluded because their balance-sheet structure and leverage are not directly comparable with those of nonfinancial firms. The resulting linked nonfinancial panel contains 35,445 firm-months across 526 firms. Depending on outcome and the required ex ante exposure, the primary regression samples contain 29,792--30,072 firm-months across 428--430 firms. Appendix Table~\ref{tab:sample_attrition} provides a compact sample-flow reconciliation and a descriptive audit of coverage near membership-spell exits. The linked sources become substantially less complete near some exits, especially M\&A-related exits, so the regression estimand should be understood as applying to the observable linked constituent sample rather than to all firms approaching index exit.

Daily market data and annual accounting variables are obtained from Compustat \citep{compustat}. We use the daily market data to construct price returns, realized volatility, and the close-high-low spread measure. Historical identity and corporate-action checks are imposed before monthly variables are calculated. The primary return is a corporate-action-consistent \emph{price} return; a common-sample specification using Yahoo adjusted total returns \citep{yahoo2026adjusted} is reported as a robustness check. ESG variables come from monthly MSCI snapshots \citep{esg}.

The timing rules are deliberately conservative but simple. For outcome month $t$, ESG must be observed in the exact preceding monthly MSCI snapshot; missing observations are not carried forward indefinitely. Annual accounting information is treated as available only after a six-month lag from the fiscal statement date. Market beta, prior volatility, prior liquidity, and momentum use information strictly preceding the outcome month. These rules make the empirical design point-in-time and prevent contemporaneous outcomes from entering the explanatory variables.

\begin{table}[htbp]
\centering
\caption{Summary statistics for the three-channel common sample.}
\label{tab:summary_stats}
\small
\begin{tabular}{lrrrrr}
\toprule
Variable & Mean & SD & P25 & Median & P75 \\
\midrule
Loss $=-$ monthly price return & -0.010 & 0.092 & -0.062 & -0.009 & 0.044 \\
Log realized volatility & -1.328 & 0.467 & -1.651 & -1.363 & -1.056 \\
Log CHL spread & -5.087 & 0.523 & -5.416 & -5.101 & -4.782 \\
MSCI Industry Adjusted ESG & 5.283 & 2.667 & 3.900 & 5.800 & 7.300 \\
Log assets & 10.045 & 1.112 & 9.267 & 9.982 & 10.788 \\
Total leverage & 0.362 & 0.188 & 0.239 & 0.351 & 0.459 \\
Cash / assets & 0.110 & 0.121 & 0.025 & 0.068 & 0.147 \\
Profitability & 0.074 & 0.070 & 0.031 & 0.063 & 0.111 \\
Investment & 0.037 & 0.032 & 0.014 & 0.028 & 0.053 \\
Momentum ($t-12$ to $t-2$) & 0.093 & 0.292 & -0.100 & 0.064 & 0.249 \\
24-month beta & 0.925 & 0.372 & 0.688 & 0.923 & 1.146 \\
Log prior RVOL12 & -1.202 & 0.350 & -1.459 & -1.236 & -0.965 \\
Prior log CHL12 & -5.095 & 0.313 & -5.319 & -5.119 & -4.898 \\
\bottomrule
\end{tabular}
\begin{minipage}{0.96\linewidth}
\footnotesize
\textit{Notes:} Statistics use the 29,792 firm-month observations in the fully risk-adjusted Joint80 sample. Accounting ratios and momentum follow the prespecified 1st/99th percentile treatment; outcomes and ESG are not winsorized.
\end{minipage}
\end{table}

\subsection{ESG and conventional firm characteristics}

The primary ESG variable, $ESG_{i,t-1}$, is MSCI's Industry Adjusted Score. MSCI's published methodology places the Weighted Average Key Issue Score (WAKIS) upstream in the scoring hierarchy and constructs IAS by normalizing that score relative to the ESG Rating Industry peer group \citep{msci2024methodology}. Historical issuer guidance further describes the weighted score as dynamic as underlying inputs change, while IAS is relatively static and corresponds to the rating-action date \citep{msci2022issuerfaq}. We use IAS as the primary construct because it aligns directly with the paper's question: whether financially material, peer-relative ESG information adds to conventional firm-level risk after broad market and sector conditions are accounted for. This choice defines the ESG construct examined; it is not based on which score produces more favorable coefficients and does not imply that IAS is objectively superior. We retain the raw score in the principal regressions: it is neither standardized nor winsorized. The pooled standard deviation in the three-channel common sample is 2.67 points. WAKIS is treated as a distinct upstream score field rather than an interchangeable version of IAS. In the post-primary paired comparison, both fields are restricted to the identical Joint80 sample and standardized separately using their standard deviations on that common sample.

Because the paired comparison becomes economically important, we validate both fields directly against the raw monthly MSCI source. Both accepted score series reproduce without unexplained mismatch, and every paired observation uses the same exact prior-month snapshot, selected raw row, and mapped entity/security. The fields nevertheless differ materially in observed update cadence, consistent with the provider's distinction between a more dynamic weighted score and a comparatively static IAS; Appendix Table~\ref{tab:measurement} reports the underlying diagnostics. Because construction and updating differ jointly, the paired comparison should be interpreted as a contrast between two stages of the provider's scoring architecture rather than as causal identification of peer normalization alone. Throughout the paper, we use \emph{score architecture} as shorthand for the combined construction and updating process represented by the two fields.

The common control vector $X_{i,t-1}$ contains six standard firm characteristics: log total assets, total leverage, cash-to-assets, profitability, investment, and momentum. Total leverage is total debt divided by assets; profitability is income before extraordinary items divided by assets; investment is capital expenditure divided by assets; and momentum is the compounded price return from months $t-12$ through $t-2$, excluding the most recent month \citep{jegadeesh1993returns}. Accounting ratios and momentum are winsorized at the prespecified pooled 1st and 99th percentile thresholds. Size, outcomes, and ESG are not winsorized.

We separately construct conventional ex ante exposure variables for each financial channel. The return model uses a 24-month daily market beta, estimated only from data preceding month $t$. The volatility model uses trailing 12-month realized volatility, while the liquidity model uses trailing 12-month log CHL spread. The joint model includes all three exposure blocks simultaneously. A 12-month beta and a 24-month downside beta are reserved for robustness and stronger return-risk adjustment, respectively.

\subsection{Financial fragility outcomes}

We use three monthly outcomes with a common sign convention: larger values represent worse financial conditions. The return outcome is monthly loss,
\begin{equation}
    Loss_{i,t}=-R_{i,t},
\end{equation}
where $R_{i,t}$ is the corporate-action-consistent monthly price return. Realized volatility is the annualized standard deviation of cleaned daily returns within month $t$,
\begin{equation}
    RVOL_{i,t}=\sqrt{252}\,\mathrm{sd}(r_{i,d}),
\end{equation}
and the regression uses $\log(RVOL_{i,t})$.

Trading cost is measured with the close-high-low bid-ask spread estimator of \citet{abdi2017spread}, which is designed for settings without intraday quote data. The estimator uses daily close, high, and low prices and has been shown to track effective-spread benchmarks well, particularly for less-liquid stocks. We aggregate valid within-month estimates and use the logarithm of the monthly CHL spread. For a targeted joint-state robustness check, we also use the already-constructed Amihud illiquidity measure \citep{amihud2002illiquidity}, which relies on return magnitude relative to trading volume and therefore supplies a conceptually different liquidity leg. CHL remains the primary liquidity measure throughout the paper.

\subsection{Joint multidimensional fragility}

The central secondary outcome captures simultaneous rather than isolated adverse states. At the feature-construction stage, before ESG coefficients were inspected, we froze pooled unconditional 80th- and 90th-percentile thresholds and their support for the three common-sample outcomes. The 80th-percentile cutoffs are a monthly loss of 5.77\%, log realized volatility of $-0.973$ (approximately 37.8\% annualized volatility), and log CHL spread of $-4.698$ (approximately a 0.91\% spread). Define
\begin{equation}
J_{i,t}^{80}
=
\mathbf{1}\{Loss_{i,t}\ge q^{L}_{0.80}\}
\mathbf{1}\{\log RVOL_{i,t}\ge q^{V}_{0.80}\}
\mathbf{1}\{\log CHL_{i,t}\ge q^{C}_{0.80}\}.
\label{eq:joint80}
\end{equation}
Thus, $J_{i,t}^{80}=1$ only when all three adverse conditions occur in the same firm-month. The indicator has 1,335 cases in the pre-regression common support audit and a 4.39\% prevalence in the fully risk-adjusted 29,792-observation sample. The more severe 90th-percentile all-three state has substantially fewer events and is correspondingly less precise.

The purpose of $J^{80}$ is not to create an ordinal ESG score or a weighted composite risk index. It preserves the economic interpretation of each underlying dimension and asks whether vulnerabilities cluster in the same firm at the same time. The one-dimensional and pairwise decompositions reported later are exploratory diagnostics of this CHL-based joint state rather than alternative primary outcomes. The Amihud robustness changes only the contemporaneous liquidity leg while leaving the loss and volatility thresholds, sample, controls, fixed effects, and ex ante exposure blocks unchanged.
\section{Empirical Design}
\label{sec:method}

\subsection{Continuous market states and marginal channels}

We model market conditions continuously rather than defining a crisis with an estimated cutoff. Let $R_t^M$ denote the monthly S\&P 500 price-index return, and define
\begin{equation}
Down_t=\max(0,-R_t^M), \qquad Up_t=\max(0,R_t^M).
\label{eq:states}
\end{equation}
The sample contains 27 negative-market months and 56 positive-market months. This parameterization retains the magnitude of the market movement rather than reducing each month to a binary regime and allows a direct distinction between directional asymmetry and a more general response to market-movement intensity.

For each marginal outcome $Y_{i,t}^{k}\in\{Loss,\log RVOL,\log CHL\}$, the fully risk-adjusted specification is
\begin{equation}
\begin{split}
Y_{i,t}^{k}
=&\;\alpha_i+\lambda_t+\beta_k ESG_{i,t-1}
+\theta_{D,k}(ESG_{i,t-1}\times Down_t)
+\theta_{U,k}(ESG_{i,t-1}\times Up_t)\\
&+\gamma_k'X_{i,t-1}+\rho_k'Z^k_{i,t-1}+\varepsilon_{i,t}^{k},
\end{split}
\label{eq:primary}
\end{equation}
where $\alpha_i$ and $\lambda_t$ are firm and month fixed effects, $X_{i,t-1}$ is the common six-variable control vector, and $Z^k_{i,t-1}$ is the channel-specific ex ante exposure block. For returns, $Z^R$ contains Beta24 and its interactions with $Down$ and $Up$; for volatility it contains prior RVOL12 and its state interactions; and for liquidity it contains prior CHL12 and its state interactions. Because $Down_t$ and $Up_t$ are common to all firms within a month, their standalone main effects are absorbed by month fixed effects.

The primary state coefficient is $\theta_{D,k}$. Because all outcomes are coded so that larger values are worse, $\theta_{D,k}<0$ is consistent with lower adverse sensitivity as the market falls. We do not infer directional asymmetry from separate significance. For every channel we directly test
\begin{equation}
H_0:\theta_{D,k}-\theta_{U,k}=0.
\label{eq:asymmetry}
\end{equation}
This test distinguishes a genuinely negative-market-specific association from a relationship that is similar across positive and negative market movements.

\subsection{Joint fragility and analytical hierarchy}

We estimate the same fixed-effects architecture with $J_{i,t}^{80}$ as the dependent variable using a linear probability model. The fully adjusted joint model includes ESG, the two ESG state interactions, the common controls, and all three conventional exposure blocks: beta, prior volatility, and prior CHL, each interacted with $Down$ and $Up$:
\begin{equation}
\begin{split}
J_{i,t}^{80}
=&\;\alpha_i+\lambda_t+\beta_J ESG_{i,t-1}
+\theta_{D,J}(ESG_{i,t-1}\times Down_t)
+\theta_{U,J}(ESG_{i,t-1}\times Up_t)\\
&+\gamma_J'X_{i,t-1}+\rho_R'Z^R_{i,t-1}+\rho_V'Z^V_{i,t-1}
+\rho_C'Z^C_{i,t-1}+u_{i,t}.
\end{split}
\label{eq:joint}
\end{equation}
The linear probability model keeps the coefficient interpretation transparent in probability points while preserving the fixed-effects and clustering architecture used for the continuous outcomes. Because the market state varies over time, the ESG state interaction can be identified from persistent cross-sectional ESG differences as well as within-firm ESG changes. The coefficient therefore captures differential adverse-state exposure associated with ESG characteristics rather than the effect of changing a firm's ESG score.

The analytical ordering is explicit. The J80 and J90 definitions and event support were frozen before ESG estimation. The primary marginal models in Equation~\eqref{eq:primary} were then estimated without any joint-outcome ESG regression. Before the first joint-outcome ESG model was estimated, J80 was designated as the main secondary joint state because its already-computed all-three support was materially larger than J90 (1,335 versus 549 events); J90 was retained as a severity check. The choice between the two frozen joint states was therefore based on pre-result event support rather than on a Joint80 ESG coefficient. The state decomposition, threshold grid, and paired ESG-architecture comparison described below were undertaken only after the primary and secondary evidence had been frozen and are explicitly treated as post-primary exploratory analyses.

\subsection{Conventional-risk benchmark and targeted validation}

To show how ordinary financial risk changes the ESG interpretation, we hold the Joint80 sample fixed and compare the full model with otherwise identical specifications that omit all three ex ante exposure blocks or add beta, prior volatility, or prior liquidity separately. These coefficient movements are descriptive attenuation, not mediation estimates.

We then use two targeted construct validations. First, we re-estimate the fully adjusted CHL-based Joint80 model with firm fixed effects and sector-by-month fixed effects on the identical 29,792-observation sample. This absorbs contemporaneous sector conditions more aggressively than ordinary month fixed effects while preserving two-way clustering by firm and calendar month. Second, we replace only the contemporaneous CHL liquidity leg of Joint80 with the already-constructed monthly Amihud illiquidity measure. The loss and RVOL thresholds, controls, ex ante exposure blocks, firm and month fixed effects, and covariance estimator remain unchanged. A single pooled 80th-percentile Amihud threshold is fixed on the common sample. No other liquidity proxy is searched.

\subsection{Post-primary localization of the clustered state}

We use two exploratory diagnostics to understand where the CHL-based association resides. The first constructs the three individual 80th-percentile tail indicators and their three pairwise intersections on the exact Joint80 sample. Each binary outcome uses the full joint architecture in Equation~\eqref{eq:joint}. The resulting fragility ladder is descriptive across outcomes; we do not rank coefficients across equations by comparing separate significance levels.

The second diagnostic decomposes the inclusive Return+RVOL state into mutually exclusive components. Define
\begin{equation}
RVOnly_{i,t}
=
\mathbf{1}\{Loss_{i,t}\ge q^L_{0.80}\}
\mathbf{1}\{\log RVOL_{i,t}\ge q^V_{0.80}\}
\mathbf{1}\{\log CHL_{i,t}<q^C_{0.80}\}.
\label{eq:rvonly}
\end{equation}
Then the inclusive Return+RVOL indicator satisfies the exact identity
\begin{equation}
PairRV80_{i,t}=RVOnly_{i,t}+J_{i,t}^{80}.
\label{eq:pairdecomp}
\end{equation}
Because all three equations use the identical sample and regressor matrix, OLS linearity implies the same identity for every estimated coefficient. We estimate $RVOnly$ and Joint80 separately and also stack the two equations with outcome-specific coefficients, firm fixed effects, and month fixed effects to test $H_0:\theta_D^{Joint80}-\theta_D^{RVOnly}=0$ under the same two-way clustered covariance. The decomposition locates the inclusive coefficient; it is not interpreted as a causal liquidity mechanism.

As an additional threshold-sensitivity diagnostic, we reconstruct the same CHL-based all-three state at the prespecified exploratory grid $q\in\{0.70,0.75,0.80,0.85,0.90\}$ on the exact Joint80 estimation sample. Every threshold uses the same fully adjusted right-hand side. The exercise is descriptive: the grid is reported in full, no threshold is selected from the results, and no monotonic severity relation is imposed.

\subsection{Paired standardized ESG-score comparison}

The final post-primary analysis asks whether MSCI's downstream IAS and upstream WAKIS encode the same relationship with Joint80. We restrict the data to the exact 29,572 firm-month observations for which IAS, WAKIS, Joint80, and all model inputs are jointly observed. Each ESG score is standardized separately using its sample standard deviation on this identical common sample. We then estimate Equation~\eqref{eq:joint} separately for standardized IAS and standardized WAKIS, preserving the same outcome, controls, exposure blocks, firm fixed effects, month fixed effects, and sample.

To estimate the difference between score constructions without conditioning one score on the other, we use a block-diagonal stacked system. The data are duplicated by score construction, and every coefficient, including all nuisance terms and fixed effects, is score specific. Both copies of a firm-month retain the same firm and month cluster identifiers. The stacked estimates reproduce the two separate equations and provide the cross-equation covariance required for paired contrasts.

Let $(\beta_I,\theta_{D,I})$ and $(\beta_W,\theta_{D,W})$ denote the standardized IAS and WAKIS coefficients. We report the supporting interaction contrast
\begin{equation}
\Delta_D=\theta_{D,I}-\theta_{D,W},
\end{equation}
and the complete standardized slope contrasts at the three market states fixed before this targeted comparison:
\begin{equation}
\Delta_0=\beta_I-\beta_W,
\end{equation}
\begin{equation}
\Delta_{-5}=(\beta_I+0.05\theta_{D,I})-(\beta_W+0.05\theta_{D,W}),
\end{equation}
\begin{equation}
\Delta_{-10}=(\beta_I+0.10\theta_{D,I})-(\beta_W+0.10\theta_{D,W}).
\label{eq:architecture_contrasts}
\end{equation}
The three complete-slope contrasts form one exploratory Holm family fixed before estimating the targeted architecture contrasts. This comparison is designed to test differences between the two score fields directly; it does not imply that either field is a superior measure of ESG. Because the supporting interaction contrast was also estimated in the same exploratory exercise, we additionally report a conservative sensitivity that Holm-adjusts all four architecture restrictions together.

\subsection{Inference, multiplicity, and economic scale}

All principal regressions include firm and month fixed effects. The baseline covariance estimator clusters standard errors two ways by firm and calendar month following standard cluster-robust practice \citep{cameron2015practitioner}. The primary multiplicity family contains exactly the three ESG $\times$ Down tests from the fully risk-adjusted return, volatility, and liquidity models; we apply Holm adjustment \citep{holm1979simple} to their two-sided $p$-values. The three primary Down-versus-Up contrasts form a separate Holm family. Exploratory component-tail, pairwise, transition, threshold, and architecture tests are kept separate from the primary adjustment.

Because the state variables vary at the monthly level and the sample contains 83 calendar months, we additionally use a wild-cluster-bootstrap procedure with the exact OLS specifications, two-way firm and month clustering, Rademacher weights, the null-imposed FNW11 procedure, and 9,999 replications; the bootstrap clustering dimension is the smaller month dimension. The three primary marginal ESG $\times$ Down hypotheses are bootstrapped and Holm-adjusted as a separate three-test family. We also report diagnostics for baseline Joint80 ESG $\times$ Down, the Joint80 Down-minus-Up contrast, Entry ESG $\times$ Down, and five complete Joint80 slopes at market returns of $-10\%$, $-5\%$, 0\%, $+5\%$, and $+10\%$. This procedure follows the logic of wild-cluster inference when conventional large-cluster approximations may be imperfect \citep{mackinnon2018wild,roodman2019fast}. Conventional and bootstrap results are reported separately; discrepancies are treated as evidence about inferential stability rather than as competing coefficient estimates.

The paired standardized architecture comparison uses the same wild-cluster design on the validated stacked system so that the same firm and month bootstrap weights apply to both architecture equations. Exactly four paired restrictions are tested: $\Delta_D$, $\Delta_0$, $\Delta_{-5}$, and $\Delta_{-10}$. The designated architecture family Holm-adjusts the three complete-slope contrasts $\{\Delta_0,\Delta_{-5},\Delta_{-10}\}$ and treats $\Delta_D$ as a supporting contrast. We also report the conservative all-four Holm sensitivity described above; the two adjustments are presented as alternative exploratory family definitions rather than as interchangeable confirmatory tests.

Economic interpretation distinguishes interaction coefficients from complete state-specific ESG slopes. For the binary Joint80 model, the complete ESG association at a market decline of magnitude $d$ is $\beta_J+d\theta_{D,J}$. We evaluate this linear combination at selected market states and translate it using the relevant ESG standard deviation into descriptive probability-point differences. These quantities are model-implied associations, not causal effects. For log outcomes, any scenario scaling remains a descriptive interaction contribution converted using $100[\exp(\Delta)-1]$.

We then conducted a targeted validation set without changing the manuscript hypotheses. The validation set comprises raw-source verification of both ESG fields and market-state support; one observed-change timing synchronization on the exact paired sample because the provider-native rating-date field does not span the full period; one 2020--2021 exclusion that retains the original Joint80 thresholds and full-sample score standardization; a GICS sector-by-month fixed-effects specification on the exact paired sample; and leave-one-month-out influence estimates for $\Delta_D$ and $\Delta_{-5}$. The timing diagnostic retains the same four paired restrictions and wild-cluster protocol, uses no future information, and searches no alternative synchronization rule. The 2020--2021 exclusion likewise searches no alternative window, and the influence exercise does not generate omission-specific bootstrap tests. These exercises assess construct validity, timing sensitivity, state support, sector-time sensitivity, period sensitivity, and single-month influence rather than create new confirmatory hypotheses.

We also retain two additional exploratory diagnostics. The Down and Up interactions can be reparameterized algebraically into market-movement-magnitude and directional components because $|R_t^M|=Down_t+Up_t$ and $R_t^M=Up_t-Down_t$; no new regression is estimated. Consecutive-month Entry and Persistence models provide a secondary dynamic perspective. Both analyses remain interpretive and do not enter the paper's central hypothesis families.
\section{Empirical Results}
\label{sec:results}

\subsection{Primary marginal evidence}

Table~\ref{tab:primary_channels} reports the fully risk-adjusted marginal specifications. The three Industry Adjusted ESG $\times$ Down coefficients are negative, but their precision differs substantially. In the return-loss model, the coefficient is $-0.0186$ with a raw $p$-value of 0.243 and a Holm-adjusted $p$-value of 0.420. The direct Down-minus-Up contrast is also imprecise ($p=0.156$). Once conventional beta exposure is modeled directly, the data do not establish a distinct ESG-return protection effect in negative market months.

The volatility channel provides the clearest primary pattern under the baseline covariance estimator. ESG $\times$ Down is $-0.0530$ with a raw $p$-value of 0.0287 and a Holm-adjusted $p$-value of 0.086. ESG $\times$ Up is also negative ($-0.0586$, raw $p=0.038$), however, and the direct Down-minus-Up contrast is far from significant ($p=0.767$). The liquidity result is again negative but imprecise: ESG $\times$ Down equals $-0.0288$ (raw $p=0.210$, Holm $p=0.420$), while the Down-minus-Up contrast has $p=0.430$.

Applying the same finite-time-cluster bootstrap protocol to the three formally primary Down interactions yields raw $p$-values of 0.436 for Return, 0.081 for RVOL, and 0.232 for CHL. Holm adjustment within this separate bootstrap family gives 0.464, 0.244, and 0.464, respectively. Thus, none of the three primary marginal hypotheses rejects even at 10\% after family adjustment under either inferential approach. Table~\ref{tab:primary_channels} reports both primary-family adjustments, while Appendix Table~\ref{tab:bootstrap_inference} gives the full side-by-side inference diagnostics.

\begin{table}[htbp]
\centering
\caption{ESG and marginal financial fragility.}
\label{tab:primary_channels}
\small
\begin{tabular}{lccc}
\toprule
 & Return loss & Log RVOL & Log CHL spread \\
\midrule
ESG & 0.0005 (0.0005) & 0.0016 (0.0020) & 0.0016 (0.0017) \\
ESG $\times$ Down & -0.0186 (0.0159) & -0.0530 (0.0242) & -0.0288 (0.0230) \\
ESG $\times$ Up & -0.0003 (0.0092) & -0.0586 (0.0283) & -0.0512 (0.0332) \\
\midrule
Conventional Holm $p$: ESG $\times$ Down & 0.420 & 0.086 & 0.420 \\
Wild-cluster Holm $p$: ESG $\times$ Down & 0.464 & 0.244 & 0.464 \\
$p$: Down $-$ Up & 0.156 & 0.767 & 0.430 \\
\midrule
Observations & 29,792 & 30,071 & 30,072 \\
Firms & 428 & 430 & 430 \\
Firm FE & Yes & Yes & Yes \\
Month FE & Yes & Yes & Yes \\
C6 controls & Yes & Yes & Yes \\
Channel-specific ex ante exposure & Beta24 & RVOL12 & CHL12 \\
\bottomrule
\end{tabular}
\begin{minipage}{0.97\textwidth}
\footnotesize
\textit{Notes:} All columns are fully risk adjusted and include firm and month fixed effects with two-way clustering by firm and month. The primary family contains the three ESG $\times$ Down tests. Conventional and wild-cluster $p$-values are Holm-adjusted separately; the wild procedure follows Section~\ref{sec:method} and Appendix Table~\ref{tab:bootstrap_inference}. Parentheses contain conventional clustered standard errors. No significance stars are used because inference is evaluated at the family level.
\end{minipage}
\end{table}

Taken together, the marginal family does not support broad, uniquely downside-specific ESG insurance. The lack of directional asymmetry is itself informative. An exploratory exact reparameterization of the estimated Down and Up coefficients yields negative market-movement-magnitude components in all four principal models, but this post-primary transformation does not replace the original hypothesis family and is not used as a new significance claim.

\subsection{Conventional risk changes the apparent ESG-resilience relation}

\begin{figure}[htbp]
\centering
\includegraphics[width=0.95\linewidth]{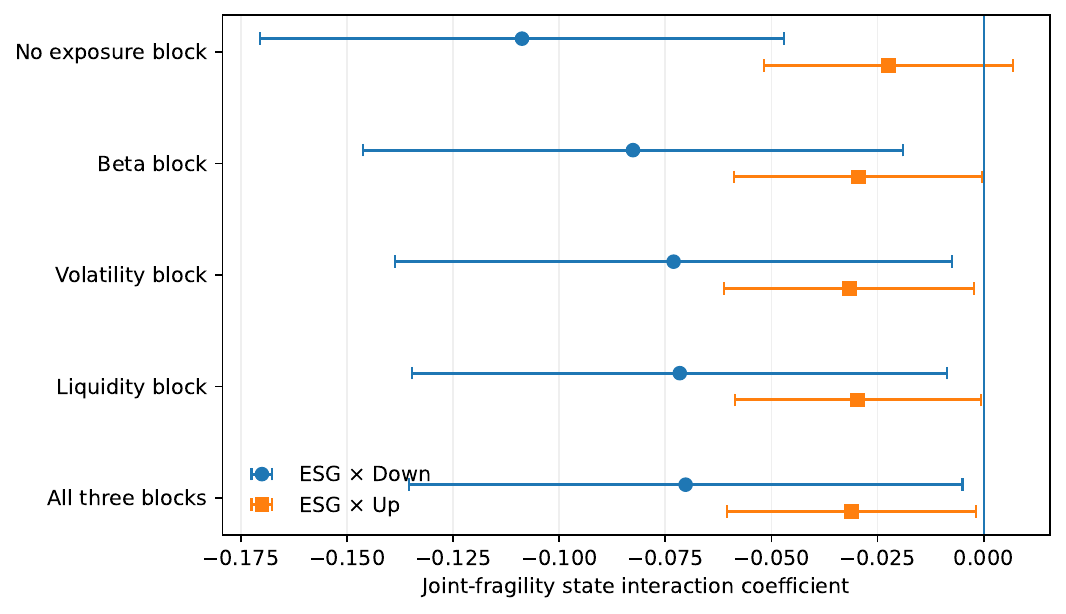}
\caption{Conventional risk adjustment and apparent state asymmetry. The figure reports CHL-based Joint80 Industry Adjusted ESG interactions with negative and positive market returns on an identical sample. ``Beta,'' ``Volatility,'' and ``Liquidity'' add the corresponding ex ante exposure block separately; ``All three blocks'' is the fully adjusted model. Error bars are 95\% two-way clustered confidence intervals. Coefficient attenuation is descriptive and is not interpreted as mediation.}
\label{fig:risk_adjustment}
\end{figure}

Figure~\ref{fig:risk_adjustment} holds the CHL-based Joint80 sample fixed and shows how the state interactions change as conventional ex ante risk blocks enter. Without beta, prior volatility, or prior liquidity, ESG $\times$ Down is $-0.1088$ ($p<0.001$), ESG $\times$ Up is $-0.0225$, and their difference is significant ($p=0.0028$). A simple specification therefore looks strongly consistent with downside-specific ESG insurance.

That interpretation weakens materially once conventional exposure is made explicit. Adding beta alone reduces the Down coefficient to $-0.0826$; prior volatility alone yields $-0.0730$; and prior liquidity alone yields $-0.0716$. With all three blocks, the coefficient is $-0.0702$ and the Down-minus-Up contrast rises to $p=0.142$. Relative to the same-sample model without the exposure blocks, the full specification attenuates the Down coefficient by about 35\%. Prior liquidity generates the largest individual attenuation of the Down coefficient, while prior volatility generates the largest attenuation of the absolute Down-versus-Up difference. These are descriptive coefficient movements, not mediation shares.

This comparison is the first central result of the paper. The estimated ESG association is materially smaller once conventional financial risk structure is modeled than under a less demanding specification that would appear strongly consistent with downside resilience. Full adjustment does not erase the negative Joint80 point estimate, motivating the question of where the residual Industry Adjusted ESG association is located.

\subsection{Where the residual Industry Adjusted ESG association resides}

The main secondary joint result uses the designated CHL-based Joint80 state. In the fully risk-adjusted model, ESG $\times$ Down equals $-0.0702$ and ESG $\times$ Up equals $-0.0311$. The conventional two-way clustered $p$-values are 0.0349 and 0.0380, respectively, while the direct Down-minus-Up contrast is not significant ($p=0.142$). The joint evidence therefore does not establish stress amplification. It documents a negative point-estimate association between Industry Adjusted ESG and the probability that adverse return, volatility, and CHL trading-cost conditions occur simultaneously.

The complete fitted ESG slope gives the economic shape of this model more directly than the interaction coefficient alone. It shifts from $0.00144$ per ESG point at a zero market return to $-0.00207$ at a 5\% decline and $-0.00558$ at a 10\% decline. Scaled by the 2.67-point sample ESG standard deviation, those slopes correspond to approximately $+0.39$, $-0.55$, and $-1.49$ probability percentage points. The fitted association therefore becomes more negative as market losses deepen. Finite-cluster inference is weaker, however: the corresponding wild-bootstrap $p$-values at 0\%, $-5\%$, and $-10\%$ are 0.174, 0.196, and 0.178. We therefore use the slope profile as economic shape rather than as evidence that any individual state-specific slope is established under the finite-cluster procedure. Appendix Table~\ref{tab:conditional_slopes} reports the full five-state diagnostic.

A post-primary decomposition clarifies why the previously estimated Return+RVOL coefficient is nearly identical to Joint80. The inclusive Return+RVOL indicator contains 1,963 events and satisfies the exact outcome identity in Equation~\eqref{eq:pairdecomp}: 656 Return+RVOL-only events plus 1,307 Joint80 events. Because all equations use the identical 29,792 observations and regressor matrix, the corresponding OLS coefficients also add exactly. Table~\ref{tab:state_localization} shows that the inclusive ESG $\times$ Down coefficient of $-0.0705$ decomposes into $-0.0003$ for Return+RVOL-only and $-0.0702$ for Joint80. Thus, essentially the entire negative coefficient on the inclusive pairwise state is allocated to observations that also satisfy the CHL tail condition.

The stacked direct contrast between the two mutually exclusive component coefficients is $-0.0699$ with $p=0.0957$. That comparison is suggestive but imprecise, so the decomposition should not be read as proof that liquidity independently causes the difference. Its contribution is more specific: it identifies where the inclusive Return+RVOL association resides and provides a stronger economic justification for treating the all-three state as a distinct clustered-fragility object.

\begin{table}[htbp]
\centering
\caption{Localization of the Return+RVOL association within mutually exclusive states.}
\label{tab:state_localization}
\small
\begin{tabular}{lrrrrrr}
\toprule
Outcome & Events & Prevalence & ESG $\times$ Down & SE & Raw $p$ & $p$(Down$-$Up) \\
\midrule
Inclusive Return+RVOL & 1,963 & 6.59\% & -0.0705 & 0.0335 & 0.035 & 0.042 \\
Return+RVOL only & 656 & 2.20\% & -0.0003 & 0.0183 & 0.986 & 0.317 \\
Joint80 & 1,307 & 4.39\% & -0.0702 & 0.0333 & 0.035 & 0.142 \\
\midrule
Joint80 $-$ Return+RVOL only & -- & -- & -0.0699 & 0.0419 & 0.096 & -- \\
\bottomrule
\end{tabular}
\begin{minipage}{0.97\textwidth}
\footnotesize
\textit{Notes:} All rows use the same 29,792 observations and regressor matrix. Return+RVOL is the exact sum of the mutually exclusive Return+RVOL-only and Joint80 indicators, so the OLS coefficients satisfy the same identity. The final row reports the direct two-way-clustered component contrast from the stacked system. The decomposition localizes the inclusive coefficient and is not interpreted as a causal liquidity mechanism.
\end{minipage}
\end{table}

The broader fragility ladder remains informative but secondary. Isolated return, RVOL, and CHL tail models are weak, while Return+RVOL has the largest pairwise negative point estimate. Appendix Table~\ref{tab:fragility_ladder} reports the full one-, two-, and three-dimensional decomposition without using separate significance levels to rank equations.

\subsection{Paired standardized ESG-score comparison}

The next result asks whether the financial relation identified with the downstream IAS is recovered using MSCI's upstream Weighted Average Key Issue Score. On the exact 29,572-observation common Joint80 sample, replacing IAS with WAKIS reverses the raw Down-state association: ESG $\times$ Down is $-0.0716$ for IAS and $+0.0341$ for WAKIS. Because the two scores have different scales, those raw coefficients are descriptive rather than directly comparable.

We therefore standardize each score using its standard deviation on the identical common sample and test the score constructions directly through the paired stacked system described in Section~\ref{sec:method}. The supporting standardized Down-interaction difference, IAS minus WAKIS, is $-0.2536$ with paired wild-cluster $p=0.0277$. More economically, at a 5\% market decline the complete one-standard-deviation IAS slope is $-0.544$ probability percentage points while the WAKIS slope is $+0.461$, a paired difference of $-1.005$ percentage points. Table~\ref{tab:architecture_comparison} reports the corresponding zero-return and 10\% decline contrasts.

The paired wild-bootstrap $p$-values for the 5\% and 10\% decline contrasts are 0.0164 and 0.0180. Both survive the three-state complete-slope Holm family fixed before estimating the targeted architecture contrasts ($p_{Holm}=0.0492$ for each). A more conservative sensitivity that treats the supporting Down-interaction contrast as part of the same four-test exploratory family yields $p_{Holm}=0.0656$ for both adverse-state contrasts. The inferential conclusion is therefore family-sensitive rather than unqualified. Market-state support also differs: nine of the 83 sample months have market returns at or below $-5\%$, placing that evaluation in the observed lower tail, whereas only March 2020 is at or below $-10\%$. We therefore treat the 5\% decline as the better-supported adverse scenario and the 10\% decline as an edge-of-support stress illustration. This post-primary comparison is not added to the original confirmatory families.

\begin{table}[htbp]
\centering
\caption{Paired standardized ESG-architecture comparison in the Joint80 model.}
\label{tab:architecture_comparison}
\scriptsize
\begin{tabular}{lrrrrrr}
\toprule
Market state & IAS slope & Weighted slope & IAS $-$ Weighted & Wild $p$ & 3-state Holm & 4-test Holm \\
 & (pp per 1 SD) & (pp per 1 SD) & (pp) & & & \\
\midrule
0\% market return & 0.412 & 0.149 & 0.263 & 0.374 & 0.374 & 0.374 \\
$-5$\% market return & -0.544 & 0.461 & -1.005 & 0.016 & 0.049 & 0.066 \\
$-10$\% market return & -1.501 & 0.772 & -2.273 & 0.018 & 0.049 & 0.066 \\
\bottomrule
\end{tabular}
\begin{minipage}{0.97\textwidth}
\footnotesize
\textit{Notes:} The paired comparison uses the identical 29,572-observation Joint80 sample. IAS and WAKIS are standardized separately on this common sample and estimated with the same fully risk-adjusted specification. Paired wild-cluster inference follows the protocol in Section~\ref{sec:method}. The 3-state Holm family contains $\{\Delta_0,\Delta_{-5},\Delta_{-10}\}$; the conservative 4-test family also includes $\Delta_D$. Nine sample months reach $-5\%$ or below and only one reaches $-10\%$, so the latter is an edge-of-support illustration. The comparison is post-primary exploratory and does not rank the two scores.
\end{minipage}
\end{table}

The paired result is also supported by direct construct validation. Across all 41,534 panel rows, both accepted ESG fields reproduce exactly from the raw MSCI snapshots, and every paired observation uses the same prior-month raw row and mapped entity/security. The fields nevertheless have materially different update cadence: IAS changes in 5.95\% of adjacent firm-month pairs, compared with 27.39\% for WAKIS. Across 37,607 overlapping firm-months, the overall Pearson correlation is 0.196 and the first-difference correlation is 0.005. MSCI's published methodology makes clear that the two fields differ not only in peer adjustment but also in update behavior \citep{msci2024methodology,msci2022issuerfaq}. Appendix Table~\ref{tab:measurement} reports the selected diagnostics.

This cadence difference motivates one post-primary timing diagnostic. Provider-native rating-action timing in the available historical source does not span the full 2019--2025 sample, so we preserve the exact 29,572-observation design and construct an observed-change synchronization proxy instead. WAKIS is allowed to refresh only when the observed IAS score changes, with no future information. The rule suppresses 83.5\% of raw WAKIS updates, lowers its adjacent-pair change rate to 5.33\%, and lengthens its median unchanged run from 3 to 12 months. Yet the standardized complete-slope difference at a 5\% market decline changes only from $-0.01005$ to $-0.00958$, an attenuation of 4.73\%. Its paired wild-cluster $p$-value is 0.0160; the three-state and conservative all-four Holm values are 0.0480 and 0.0640. Thus, synchronizing the observed update cadence does not account for most of the adverse-state difference. Because observed IAS score changes do not capture rating actions that leave IAS unchanged, the diagnostic does not isolate peer normalization, analyst review, or another specific provider mechanism. Appendix Table~\ref{tab:timing_sync} reports the diagnostic.

A final targeted validation asks whether the paired score contrast is absorbed by broad sector-specific monthly shocks or dominated by one calendar month. Replacing ordinary month fixed effects with GICS sector-by-month fixed effects reduces the 5\% complete-slope difference from $-1.005$ to $-0.743$ probability percentage points, a 26\% attenuation, while preserving its negative sign; the conservative all-four Holm value is 0.0488. In the accepted ordinary-month specification, all 83 leave-one-month-out estimates of $\Delta_{-5}$ remain negative. The largest attenuation is 22.3\%, and March 2020 is not the most influential month. These exercises are coefficient-preservation and influence diagnostics rather than a new significance search. Appendix Table~\ref{tab:architecture_validation} reports the full results.

\subsection{Scope, inference, and measurement limits}

Three results limit how far the preceding patterns can be generalized. First, finite-time-cluster inference is materially more conservative for the baseline Joint80 coefficient. The conventional two-way clustered $p$-value for ESG $\times$ Down is 0.0349, whereas the null-imposed wild-cluster-bootstrap $p$-value is 0.1572. The direct Down-minus-Up contrast moves from 0.1415 to 0.2531. The coefficient itself is unchanged because the bootstrap concerns inference rather than estimation. We therefore distinguish coefficient stability from inferential stability rather than classifying the Joint80 result as simply ``robust'' or ``not robust.''

Second, the CHL-based result is liquidity-measure sensitive. Amihud is observed for all 29,792 baseline observations, so the comparison is not driven by sample attrition. A pooled 80th-percentile Amihud leg produces 727 all-three events, compared with 1,307 CHL-based events, with only 532 observations shared by the two indicators. ESG $\times$ Down remains negative but falls to $-0.0254$ ($p=0.457$). We therefore interpret the primary state specifically as spread-based trading-cost fragility and do not generalize it to a proxy-invariant liquidity factor.

Third, the exploratory J70--J90 threshold profile is non-monotonic. The estimates are smaller under the broader J70 and J75 definitions and more negative across the upper-tail definitions, but only J80 is conventionally significant. Appendix Table~\ref{tab:threshold_sensitivity} reports the complete grid. The exercise therefore reduces concern that J80 is a solitary negative realization without establishing either a discrete cutoff or a monotonic severity relation.

A final period-sensitivity diagnostic removes all 2020--2021 outcome months while retaining the original Joint80 thresholds and score standardization. The baseline IAS Down interaction remains negative but has wild-cluster $p=0.290$. The paired 5\% complete-slope difference also retains its sign at $-0.00725$, with wild-cluster $p=0.155$, and no architecture contrast survives Holm adjustment. We therefore describe the pattern as directionally recognizable but inferentially weaker, not as a robustness success. No 10\% market-decline month remains after the exclusion. Appendix Table~\ref{tab:covid_sensitivity} reports the comparison.

The coefficient also shows specification and influence stability that should be interpreted separately from these inferential limits. Omitting one month at a time leaves the Joint80 Down interaction negative in all 83 estimations. Replacing month fixed effects with sector-by-month fixed effects on the identical sample yields ESG $\times$ Down of $-0.0532$, about 24\% smaller in absolute magnitude than the baseline. The role of this specification is to show that contemporaneous broad sector conditions do not account for the full coefficient; it does not reproduce MSCI's finer proprietary peer groups, and a defensible historical point-in-time SIC or NAICS classification is not available in the historical data used here.

\subsection{Additional exploratory diagnostics}

The transition analysis is retained as interpretation rather than a headline contribution. Among firms outside Joint80 in the prior month, ESG $\times$ Down is $-0.0758$ under conventional inference, but the wild-cluster-bootstrap diagnostic is 0.1086. The Persistence sample is much smaller, with 1,279 observations versus 28,082 for Entry, and no corresponding Persistence association is detected. We therefore do not claim a formal resistance-versus-recovery mechanism. Detailed transition results are reported in Appendix Table~\ref{tab:transitions}.
\section{Discussion}
\label{sec:discussion}

\subsection{Reframing the ESG-resilience benchmark}

The first implication is that ESG resilience is a more demanding empirical proposition once conventional exposure is allowed to vary with market conditions. In the simpler Joint80 specification, the negative-market interaction and the Down-versus-Up contrast resemble downside insurance. Explicitly conditioning on beta, prior volatility, and prior liquidity materially reduces that pattern, while the formally primary marginal family does not survive multiplicity adjustment under either conventional or finite-cluster inference. We interpret these changes as evidence that conventional exposures account for part of the estimated association, not as a mediation decomposition of an ESG effect.

This distinction matters because investors already observe many of the characteristics that make firms more or less exposed to weak markets. An ESG coefficient is financially more informative when it survives a benchmark that gives beta, volatility, and liquidity their own state dependence rather than forcing ESG to absorb those differences. The paper therefore separates apparent resilience from incremental association. ESG need not be financially irrelevant for a broad crisis-insurance interpretation to be too strong.

\subsection{What the clustered-state localization means}

The exact decomposition sharpens the economic interpretation of the residual Industry Adjusted ESG pattern. The inclusive Return+RVOL coefficient is almost entirely allocated to observations that also breach the CHL trading-cost tail, while the mutually exclusive Return+RVOL-only component is essentially zero. Because the decomposition uses the same observations and regressor matrix, this localization follows directly from the outcome identity rather than from comparing significance across unrelated regressions.

The result does not identify an independent liquidity mechanism. The direct component contrast is imprecise, the baseline Joint80 coefficient is substantially less precise under wild-cluster inference than under the conventional covariance estimator, and the corresponding Amihud construction is weaker. The threshold profile is also non-monotonic. These limits narrow the claim in a useful way: the evidence concerns a CHL-based clustered state in which adverse return, volatility, and trading-cost conditions occur together. It does not establish that liquidity causes ESG resilience or that a universal multidimensional fragility factor has been identified.

The complete fitted ESG slope is consistent with the same interpretation. Its point estimate becomes more negative as market losses deepen, but the state-specific slope restrictions are not individually established under the finite-cluster procedure. The fitted shape is therefore informative about the model's economic profile without supporting a renewed claim that ESG protection activates uniquely in falling markets.

\subsection{Measurement dependence as a finance result}

The paired score comparison turns ESG measurement from a routine robustness choice into part of the economic result. MSCI's downstream IAS and upstream WAKIS occupy different stages of the same scoring architecture, yet they imply materially different fitted relationships when the outcome, firm-month sample, controls, fixed effects, and estimation design are held constant. Standardizing both scores and estimating them in a paired system removes differences in scale and permits a direct test of the financial contrast. At a 5\% market decline, the fitted difference is about one probability percentage point per score standard deviation, an economically meaningful magnitude for an outcome with a relatively low unconditional incidence.

The targeted validations make several simple explanations less persuasive without converting the finding into confirmatory evidence. The adverse-state difference remains negative after absorbing broad sector-specific monthly shocks, every leave-one-month-out estimate preserves its sign, and synchronizing the observed WAKIS update cadence much more closely to IAS leaves most of the contrast intact. The comparison is nevertheless exploratory and its inferential strength depends on the multiplicity family. Removing 2020--2021 also reduces finite-cluster precision. The combined evidence is therefore stronger as a statement about coefficient stability and measurement dependence than as a claim about one universally significant stress-state effect.

MSCI's published methodology helps define what this comparison can and cannot identify. WAKIS is upstream of IAS, which applies industry-peer normalization, and the two fields also differ in update behavior \citep{msci2024methodology,msci2022issuerfaq}. The timing diagnostic indicates that observed update cadence alone does not account for most of the paired difference, but it cannot separate peer normalization, analyst review, rating actions that leave IAS unchanged, or other elements of the provider's production process. We therefore do not infer that IAS is the correct ESG measure or that WAKIS is inferior. The finance implication is more general: an empirical relation attributed to ``ESG'' can depend materially on the score construction used to represent it, even within one provider. This complements the broader evidence that scope, measurement, and weighting contribute to ESG rating divergence \citep{berg2022aggregate,christensen2022corporate,avramov2022sustainable,lou2025disagreement}.

\subsection{Scope, implications, and limitations}

Several features bound the population and states to which these findings apply. The sample contains 83 calendar months, only 27 negative-market months, and limited support for very large market declines. The 10\% downside scenario is therefore an edge-of-support illustration rather than a central empirical state. Excluding 2020--2021 preserves the direction of the baseline and architecture patterns but materially reduces finite-cluster precision. This is consistent with the leave-one-month-out evidence: no single month dominates the paired score contrast, yet the period containing the strongest aggregate market variation contributes important identifying variation.

The linked constituent design imposes a separate scope condition. Coverage deteriorates near some S\&P 500 membership-spell endings, particularly identified M\&A exits. Membership exit is not a distress classification, and the missingness does not identify a bias of known sign, but the estimates apply to the observable linked nonfinancial constituent sample rather than to all firms approaching index exit. The observational design also precludes causal interpretation. Firm and time fixed effects, point-in-time controls, and ex ante exposures reduce important confounding channels, but they do not show that changing a firm's ESG score changes its financial fragility. The within-between diagnostic provides no strong basis for such an interpretation.

The analytical hierarchy should also remain explicit. The marginal risk-adjusted tests are the formally primary family. Joint80 is a designated secondary state, while the state decomposition, threshold grid, complete-slope bootstrap, paired score comparison, and subsequent validations are post-primary analyses used to refine interpretation. Keeping those layers separate is preferable to treating the most favorable exploratory result as though it had been the original hypothesis.

For empirical ESG finance, the broader implication is methodological but economically substantive. State-dependent ESG coefficients should be benchmarked against equally state-dependent conventional exposures, and the ESG score itself should be treated as part of the estimand rather than as a neutral input. For investors, the same caution applies: a high ESG score is not automatically a generic resilience signal, because its information content depends on both the financial state being evaluated and the construction of the score. The paper therefore favors more specific claims about which ESG measure is related to which financial outcome under which market conditions, rather than a measurement-invariant notion of ESG protection.
\section{Conclusion}
\label{sec:conclusion}

This paper asks whether apparent ESG resilience remains after conventional financial exposure is modeled explicitly, where any residual association is concentrated, and whether the resulting inference is stable to the published ESG score used. Using a point-in-time panel of nonfinancial S\&P 500 constituents from 2019 to 2025, we examine return loss, realized volatility, trading cost, and a CHL-based simultaneous adverse state under a common fixed-effects framework with ex ante return, volatility, and liquidity exposures.

Three conclusions emerge. First, the broad resilience narrative weakens under the stronger financial benchmark. Conventional exposure materially attenuates the negative-market ESG association, and the formally primary marginal tests provide limited evidence after multiplicity adjustment and finite-cluster inference. Second, the remaining Industry Adjusted ESG pattern is highly localized. An exact decomposition shows that the negative coefficient on the inclusive Return+RVOL state is almost entirely allocated to observations that also cross the CHL trading-cost tail, while the mutually exclusive Return+RVOL-only component is essentially zero. This localization is economically informative but does not establish a causal liquidity mechanism and is sensitive to the liquidity measure and finite-cluster inference.

%This localization is economically informative but does not establish a causal liquidity mechanism and is sensitive to the liquidity measure and finite-cluster inference.

Third, the residual relation is measurement dependent. On an identical sample, standardized IAS and WAKIS imply materially different fitted associations in adverse market states. Sector-time conditioning, month-by-month influence analysis, and an observed-change timing synchronization leave most of the central 5\% market-decline contrast intact. These diagnostics make broad sector shocks, a single calendar month, and observed update cadence less persuasive as complete explanations, but the comparison remains post-primary exploratory and does not identify which element of MSCI's production process generates the difference.

The evidence therefore supports a narrower view of ESG resilience than a generic crisis insurance interpretation. Conventional financial exposure, the definition of the adverse state, and the construction of the ESG score materially shape the estimated association between ESG and financial fragility. These choices should therefore be treated as substantive features of the empirical design rather than as interchangeable implementation details.

%The evidence therefore supports a narrower view of ESG resilience than a generic crisis-insurance interpretation. Conventional risk, the definition of the adverse financial state, and the construction of the ESG score all affect the relation that an empirical design attributes to ESG. For finance research and investment interpretation, those elements should be treated as part of the economic object being estimated rather than as interchangeable implementation details.

\FloatBarrier
% \section*{Declaration of competing interest}
% The authors declare that they have no known competing financial interests or personal relationships that could have appeared to influence the work reported in this paper.

\section*{Data availability}
The analysis combines Compustat market and accounting data, index-membership data, and MSCI ESG data that are subject to their respective licensing restrictions and therefore cannot be redistributed by the authors. Variable definitions, empirical specifications, and the timing conventions needed to reproduce the analysis from appropriately licensed source data are described in the paper and appendix.

\FloatBarrier
\appendix
\setcounter{table}{0}
\renewcommand{\thetable}{A\arabic{table}}
\section{Additional Evidence}
\label{sec:appendix}

\subsection{Sample flow and joint-state validation}

Table~\ref{tab:sample_attrition} reconciles the linked nonfinancial constituent panel with the final model samples using only the frozen eligibility rules. The main reduction occurs when the exact prior-month ESG observation and the 11-month momentum history are required. The return, volatility, and liquidity specifications then branch from the common C6 sample according to their ex ante exposure requirements. Panel B adds the descriptive coverage audit around membership-spell ends and makes explicit the much weaker market-data coverage near identified M\&A exits.

\begin{table}[H]
\centering
\caption{Sample construction and coverage near membership-spell ends.}
\label{tab:sample_attrition}
\small
\setlength{\tabcolsep}{5.8pt}
\begin{tabularx}{\tblwidth}{@{}>{\raggedright\arraybackslash}Xrr@{}}
\toprule
\multicolumn{3}{@{}l}{\textit{Panel A: model eligibility}} \\
Sample stage & Firm-months & Firms \\
\midrule
Nonfinancial linked constituent panel & 35,445 & 526 \\
Exact prior-month Industry Adjusted ESG & 32,703 & 483 \\
Five accounting controls available & 32,630 & 482 \\
Plus momentum requirement (C6) & 30,074 & 430 \\
Return R-B eligibility & 29,792 & 428 \\
RVOL V-B eligibility & 30,071 & 430 \\
CHL L-B eligibility & 30,072 & 430 \\
Full Joint80 J-B eligibility & 29,792 & 428 \\
\end{tabularx}

\vspace{0.6em}
\begingroup
\setlength{\tabcolsep}{3.2pt}
\begin{tabularx}{\tblwidth}{@{}>{\raggedright\arraybackslash}Xrrr@{}}
\toprule
\multicolumn{4}{@{}l}{\textit{Panel B: descriptive coverage around membership-spell ends}} \\
Window / group & Market usable & Exact prior IAS & Full J-B eligible \\
\midrule
Final 1 month, all ending spells & 54.7\% & 82.9\% & 39.3\% \\
Final 1 month, identified M\&A exits & 4.7\% & 86.0\% & 4.7\% \\
Final 3 months, all ending spells & 55.7\% & 82.6\% & 39.1\% \\
Final 3 months, identified M\&A exits & 5.6\% & 85.6\% & 4.0\% \\
Non-exit observations, same calendar years & 99.0\% & 92.2\% & 75.4\% \\
\bottomrule
\end{tabularx}
\endgroup
\begin{minipage}{\tblwidth}
\footnotesize
\textit{Notes:} Panel A traces cumulative eligibility through C6 and the outcome-specific exposure branches. Panel B is a source-coverage audit, not a distress classification. Coverage is substantially weaker near some membership-spell endings, particularly identified M\&A exits; the regression estimand therefore applies to the observable linked constituent sample.
\end{minipage}
\end{table}

Table~\ref{tab:joint_validation} reports the two targeted construct validations of the CHL-based Joint80 specification. Sector-by-month fixed effects retain the exact baseline sample and leave the ESG $\times$ Down coefficient economically substantial, although smaller. The Amihud alternative also retains the exact sample but defines a substantially rarer joint state and yields a smaller, imprecise negative coefficient. The comparison is reported as liquidity-measure sensitivity rather than as evidence that either proxy is universally preferred.

\begin{table}[H]
\centering
\caption{Targeted construct validation of the joint fragility result.}
\label{tab:joint_validation}
\small
\begin{tabular}{lccc}
\toprule
 & Baseline CHL Joint80 & Sector $\times$ month FE & Amihud Joint80 \\
\midrule
ESG $\times$ Down & -0.0702 & -0.0532 & -0.0254 \\
Conventional $p$ (Down) & 0.0349 & 0.0002 & 0.4569 \\
ESG $\times$ Up & -0.0311 & -0.0275 & -0.0164 \\
Conventional $p$ (Up) & 0.0380 & 0.0236 & 0.1315 \\
$p$(Down$-$Up) & 0.1415 & 0.0338 & 0.7381 \\
Joint events & 1,307 & 1,307 & 727 \\
Prevalence & 4.39\% & 4.39\% & 2.44\% \\
Observations & 29,792 & 29,792 & 29,792 \\
Firm FE & Yes & Yes & Yes \\
Time FE & Month & Sector $\times$ month & Month \\
Full ex ante exposure blocks & Yes & Yes & Yes \\
\bottomrule
\end{tabular}
\begin{minipage}{0.97\textwidth}
\footnotesize
\textit{Notes:} The baseline uses CHL in Joint80. The sector-by-month specification retains the same outcome and sample but replaces month fixed effects with sector-by-month fixed effects. The Amihud specification changes only the contemporaneous liquidity leg, using a pooled 80th-percentile threshold on the same sample. These checks assess coefficient and measurement sensitivity; their conventional $p$-values do not replace the baseline finite-cluster inference.
\end{minipage}
\end{table}

Table~\ref{tab:conditional_slopes} reports the complete fitted Industry Adjusted ESG slope from the frozen Joint80 model at five bounded market states together with the corresponding wild-cluster diagnostics. The conventional slope profile becomes more negative as downside magnitude increases, but none of the five restrictions rejects at 5\% under the finite-cluster procedure. The table also reports the empirical support distinction underlying the 5\% and 10\% downside scenarios.

\begin{table}[H]
\centering
\caption{Complete Joint80 ESG slopes and finite-cluster diagnostics.}
\label{tab:conditional_slopes}
\small
\begin{tabular}{lrrrrr}
\toprule
Market return & ESG slope & Conventional 95\% CI & 1 ESG SD (pp) & Conventional $p$ & Wild $p$ \\
\midrule
$-10$\% & -0.00558 & [-0.01099, -0.00016] & -1.49 & 0.043 & 0.178 \\
$-5$\% & -0.00207 & [-0.00446, 0.00033] & -0.55 & 0.091 & 0.196 \\
0\% & 0.00144 & [-0.00041, 0.00330] & 0.39 & 0.127 & 0.174 \\
$+5$\% & -0.00011 & [-0.00114, 0.00092] & -0.03 & 0.833 & 0.833 \\
$+10$\% & -0.00166 & [-0.00339, 0.00006] & -0.44 & 0.059 & 0.069 \\
\bottomrule
\end{tabular}
\begin{minipage}{0.97\textwidth}
\footnotesize
\textit{Notes:} Complete ESG slopes are linear combinations of the estimated Joint80 coefficients. The one-SD column uses the 2.67-point IAS standard deviation and reports probability percentage points. Conventional intervals use two-way firm/month clustering; wild $p$-values follow the finite-cluster procedure in Section~\ref{sec:method}. None of the five slope restrictions rejects at 5\% under wild inference. Nine months reach $-5\%$ or below and only one reaches $-10\%$, placing the latter near the empirical edge. The slopes are descriptive associations, not causal effects.
\end{minipage}
\end{table}

Table~\ref{tab:bootstrap_inference} reports wild-cluster-bootstrap inference for the formally primary marginal family and the bounded Joint80/Entry diagnostics. The primary bootstrap $p$-values are Holm-adjusted within their own three-test family and reported separately from the conventional Holm family. The bootstrap leaves point estimates unchanged but generally yields more conservative inference. These diagnostics motivate the distinction between coefficient stability and inferential stability used in the main text.

\begin{table}[H]
\centering
\caption{Wild-cluster-bootstrap inference diagnostics.}
\label{tab:bootstrap_inference}
\small
\begin{tabular}{lcccc}
\toprule
Hypothesis & Two-way raw $p$ & Two-way Holm $p$ & Wild-cluster raw $p$ & Wild-cluster Holm $p$ \\
\midrule
\multicolumn{5}{l}{\textit{Panel A. Formally primary marginal family}} \\
Return: ESG $\times$ Down $=0$ & 0.2431 & 0.4195 & 0.4363 & 0.4644 \\
RVOL: ESG $\times$ Down $=0$ & 0.0287 & 0.0860 & 0.0812 & 0.2436 \\
CHL: ESG $\times$ Down $=0$ & 0.2098 & 0.4195 & 0.2322 & 0.4644 \\
\midrule
\multicolumn{5}{l}{\textit{Panel B. Targeted secondary and exploratory diagnostics}} \\
Joint80: ESG $\times$ Down $=0$ & 0.0349 & -- & 0.1572 & -- \\
Joint80: Down $-$ Up $=0$ & 0.1415 & -- & 0.2531 & -- \\
Entry: ESG $\times$ Down $=0$ & 0.0204 & 0.0408$^{a}$ & 0.1086 & -- \\
\bottomrule
\end{tabular}
\begin{minipage}{0.97\linewidth}
\footnotesize
\textit{Notes:} Conventional inference uses two-way firm/month clustered covariance. Wild inference uses the same OLS specifications with the procedure described in Section~\ref{sec:method}. Panel A reports separate three-test Holm families for conventional and wild $p$-values. Panel B reports the Joint80 and Entry diagnostics. $^{a}$The conventional Entry Holm value belongs to the separate Entry/Persistence family; only Entry was bootstrapped. The bootstrap changes inference, not point estimates.
\end{minipage}
\end{table}

\FloatBarrier
\subsection{Conventional risk adjustment and robustness}

Table~\ref{tab:risk_attenuation} reports the same-sample Joint80 specifications used to assess how conventional ex ante exposure changes the state-dependent ESG coefficients. The comparison is intentionally descriptive: the coefficient movements are not interpreted as mediation or as the share of an ESG effect explained by a particular exposure.

\begin{table}[H]
\centering
\caption{Joint fragility and conventional risk adjustment.}
\label{tab:risk_attenuation}
\small
\begin{tabular}{lrrrrr}
\toprule
 & No exposure & Beta only & Volatility only & Liquidity only & Full \\
\midrule
ESG $\times$ Down & -0.1088 & -0.0826 & -0.0730 & -0.0716 & -0.0702 \\
Raw $p$ (Down) & 0.001 & 0.011 & 0.029 & 0.026 & 0.035 \\
ESG $\times$ Up & -0.0225 & -0.0295 & -0.0317 & -0.0297 & -0.0311 \\
Raw $p$ (Up) & 0.133 & 0.048 & 0.035 & 0.045 & 0.038 \\
$p$(Down$-$Up) & 0.003 & 0.045 & 0.121 & 0.114 & 0.142 \\
\bottomrule
\end{tabular}
\begin{minipage}{0.97\textwidth}
\footnotesize
\textit{Notes:} All columns use the same 29,792-observation Joint80 sample. Restricted specifications add the indicated ex ante exposure block; the Full model includes beta, prior volatility, and prior liquidity together. Coefficient changes are descriptive attenuation, not mediation effects.
\end{minipage}
\end{table}

Table~\ref{tab:robustness} summarizes selected frozen robustness checks for the three primary marginal models. The volatility interaction remains negative in the principal sample and across most alternative specifications, although it becomes smaller and imprecise with sector-by-month fixed effects. Return and CHL estimates are generally negative but less precise. The weighted-average ESG row is reported as measurement sensitivity rather than as a near-equivalent alternative ESG measure.

\begin{table}[H]
\centering
\caption{Selected robustness checks for ESG $\times$ Down.}
\label{tab:robustness}
\scriptsize
\begin{tabular}{lccc}
\toprule
 & Return & RVOL & CHL \\
\midrule
2020 start & -0.0226 (0.163) & -0.0618 (0.016) & -0.0314 (0.178) \\
Active at Nov. 2025 & -0.0219 (0.253) & -0.0614 (0.015) & -0.0265 (0.287) \\
Sector $\times$ month FE & -0.0165 (0.160) & -0.0295 (0.210) & -0.0063 (0.809) \\
Financial-strength interactions & -0.0171 (0.189) & -0.0496 (0.039) & -0.0429 (0.047) \\
Book-to-market & -0.0174 (0.260) & -0.0661 (0.006) & -0.0428 (0.140) \\
Tangibility & -0.0196 (0.241) & -0.0512 (0.040) & -0.0289 (0.217) \\
Exclude real estate & -0.0235 (0.178) & -0.0559 (0.038) & -0.0310 (0.228) \\
Exclude 2020--2021 & -0.0115 (0.316) & -0.0680 (0.026) & -0.0559 (0.112) \\
Yahoo adjusted total return & -0.0218 (0.176) & -- & -- \\
12-month beta & -0.0196 (0.193) & -- & -- \\
24-month downside beta & -0.0189 (0.214) & -- & -- \\
Weighted-average ESG & 0.0070 (0.625) & 0.0211 (0.589) & 0.0709 (0.225) \\
\bottomrule
\end{tabular}
\begin{minipage}{0.97\textwidth}
\footnotesize
\textit{Notes:} Cells report ESG $\times$ Down coefficients with raw two-sided $p$-values in parentheses. These are prespecified robustness checks, not new multiplicity families. Return-only checks use the alternative beta construction documented in the analysis archive. ``--'' denotes not applicable. Weighted Average ESG is a measurement-sensitivity comparison rather than a near-equivalent substitute for IAS.
\end{minipage}
\end{table}

\FloatBarrier
\subsection{Exploratory state decomposition and threshold sensitivity}

Table~\ref{tab:fragility_ladder} reports the complete one-, two-, and three-dimensional CHL-based fragility ladder on the exact 29,792-observation Joint80 sample. The one-dimensional and two-dimensional families are exploratory and are Holm-adjusted separately. The table is used to describe the progression of point estimates rather than to claim a cross-equation ranking.

\begin{table}[H]
\centering
\caption{From marginal tails to multidimensional fragility.}
\label{tab:fragility_ladder}
\small
\begin{tabular}{clrrrrr}
\toprule
Dimension & Outcome & Prevalence & ESG $\times$ Down & Raw $p$ & Within-family Holm $p$ & $p$(Down$-$Up) \\
\midrule
1D & Return tail & 19.96\% & -0.0423 & 0.411 & 0.853 & 0.886 \\
1D & RVOL tail & 19.87\% & -0.0321 & 0.284 & 0.853 & 0.670 \\
1D & CHL tail & 19.87\% & 0.0153 & 0.709 & 0.853 & 0.772 \\
2D & Return + RVOL & 6.59\% & -0.0705 & 0.035 & 0.106 & 0.042 \\
2D & Return + CHL & 5.92\% & -0.0493 & 0.197 & 0.395 & 0.519 \\
2D & RVOL + CHL & 13.37\% & -0.0256 & 0.390 & 0.395 & 0.732 \\
3D & Joint80 & 4.39\% & -0.0702 & 0.035 & -- & 0.142 \\
\bottomrule
\end{tabular}
\begin{minipage}{0.97\textwidth}
\footnotesize
\textit{Notes:} All seven outcomes use the same 29,792 observations and fully risk-adjusted specification. The one- and two-dimensional families are exploratory and Holm-adjusted separately within their three-test families. Joint80 is the designated secondary all-three state and is reported as a single test. No cross-equation test ranks coefficients across rows.
\end{minipage}
\end{table}

Table~\ref{tab:threshold_sensitivity} reports the full exploratory J70--J90 grid. The profile is non-monotonic: J70 and J75 have smaller coefficients, while J80, J85, and J90 are more negative. Only J80 is conventionally significant. The exercise therefore does not support a smooth severity gradient or a uniquely privileged threshold.

\begin{table}[H]
\centering
\caption{Exploratory threshold sensitivity of the CHL-based joint state.}
\label{tab:threshold_sensitivity}
\small
\begin{tabular}{lrrrrrr}
\toprule
Joint threshold & Events & Prevalence & ESG $\times$ Down & SE & Raw $p$ & Complete slope at $-10$\% \\
\midrule
70th percentile & 2,469 & 8.29\% & -0.0236 & 0.0471 & 0.616 & -0.00200 \\
75th percentile & 1,854 & 6.22\% & -0.0205 & 0.0411 & 0.617 & -0.00130 \\
80th percentile & 1,307 & 4.39\% & -0.0702 & 0.0333 & 0.035 & -0.00558 \\
85th percentile & 873 & 2.93\% & -0.0512 & 0.0347 & 0.141 & -0.00394 \\
90th percentile & 532 & 1.79\% & -0.0690 & 0.0543 & 0.204 & -0.00534 \\
\bottomrule
\end{tabular}
\begin{minipage}{0.97\textwidth}
\footnotesize
\textit{Notes:} All thresholds use the same 29,792 observations and fully risk-adjusted right-hand side. The exploratory grid $\{70,75,80,85,90\}$ was fixed before this exercise and is reported in full. The non-monotonic profile does not establish either a discrete cutoff or a smooth severity gradient.
\end{minipage}
\end{table}

\FloatBarrier
\subsection{Exploratory transitions}

Table~\ref{tab:transitions} reports the Entry and Persistence models. The conventional Entry interaction is negative, but the wild-cluster-bootstrap diagnostic reported in Table~\ref{tab:bootstrap_inference} yields weaker inference. The Persistence sample is far smaller and no corresponding Persistence association is detected. No formal cross-model test establishes that the Entry and Persistence coefficients differ.

\begin{table}[H]
\centering
\caption{Exploratory transitions into and within joint fragility.}
\label{tab:transitions}
\small
\begin{tabular}{lcc}
\toprule
 & Entry & Persistence \\
\midrule
ESG & 0.0018 (0.0010) & 0.0003 (0.0088) \\
ESG $\times$ Down & -0.0758 (0.0327) & 0.0254 (0.0670) \\
ESG $\times$ Up & -0.0419 (0.0171) & -0.0001 (0.0979) \\
Holm $p$: ESG $\times$ Down & 0.041 & 0.705 \\
Wild-cluster $p$: ESG $\times$ Down & 0.109 & --- \\
$p$: Down $-$ Up & 0.180 & N/C \\
\midrule
Observations & 28,082 & 1,279 \\
Event rate & 4.10\% & 11.10\% \\
Firm FE & Yes & Yes \\
Month FE & Yes & Yes \\
Full ex ante exposure blocks & Yes & Yes \\
\bottomrule
\end{tabular}
\begin{minipage}{0.96\linewidth}
\footnotesize
\textit{Notes:} Entry uses observations with Joint80$_{t-1}=0$ and Persistence uses Joint80$_{t-1}=1$, requiring consecutive calendar months. The two Down interactions form a separate exploratory Holm family. The Entry wild diagnostic follows Appendix Table~\ref{tab:bootstrap_inference}; Persistence was not included in that protocol. ``N/C'' denotes a non-computable two-way-clustered contrast variance. No formal cross-model test compares Entry with Persistence.
\end{minipage}
\end{table}

\FloatBarrier
\subsection{Within-firm, pillar, and influence diagnostics}

Two additional decompositions are retained as secondary diagnostics rather than as central findings. First, the Industry Adjusted ESG score is separated into a firm-specific mean and a within-firm deviation. Across the three marginal channels, none of the six Down interactions for the within and firm-mean components survives the frozen Holm adjustment; the smallest adjusted $p$-value is 0.331. The data therefore do not identify whether the primary state pattern is principally associated with changes in a firm's own ESG score or with persistent cross-firm ESG differences.

Second, Environmental, Social, and Governance pillar scores are examined separately in exactly nine risk-adjusted channel models. None of the nine pillar $\times$ Down interactions survives its Holm family; the smallest adjusted $p$-value is 0.214. A Governance-CHL Down-versus-Up contrast survives the separate asymmetry family, but both state interactions are positive and the Up coefficient is larger. We therefore do not interpret the pillar evidence as a favorable ESG-resilience mechanism.

The primary marginal models are also re-estimated after omitting one month at a time. All 249 ESG $\times$ Down estimates remain negative. This sign stability is an influence diagnostic and is not interpreted as a substitute for statistical precision or multiplicity-adjusted inference.

\FloatBarrier
\subsection{ESG measurement architecture}

Table~\ref{tab:measurement} reports the descriptive comparison between MSCI's Industry Adjusted and Weighted Average aggregate scores together with the raw-source validation motivated by their weak dependence. Both accepted fields reproduce exactly from the raw MSCI snapshots and use the same vintage and mapped entity throughout the paired sample, while their observed update frequencies differ materially. MSCI's published hierarchy places the weighted score upstream of IAS, and historical issuer guidance describes the weighted score as more dynamic while IAS is comparatively static and linked to rating actions \citep{msci2024methodology,msci2022issuerfaq}. The main text complements these diagnostics with the paired standardized Joint80 comparison in Table~\ref{tab:architecture_comparison}. Because GICS sector is the finest frozen peer-group field available, the analysis does not establish what would occur under finer historical industry or sub-industry conditioning.

\begin{table}[H]
\centering
\caption{MSCI score-architecture diagnostics and raw-source validation.}
\label{tab:measurement}
\small
\begin{tabular}{lr}
\toprule
Diagnostic & Value \\
\midrule
\multicolumn{2}{l}{\textit{Panel A: architecture overlap and dependence}} \\
Overall Pearson correlation & 0.196 \\
Overall Spearman correlation & 0.478 \\
Within-firm demeaned correlation & 0.118 \\
Month-to-month first-difference correlation & 0.005 \\
Pooled within-sector-month Pearson & 0.166 \\
Pooled within-sector-month Spearman & 0.421 \\
Median within-sector-month Spearman & 0.499 \\
Median top-quintile overlap & 60.0\% \\
Median bottom-quintile overlap & 33.3\% \\
Share of IAS-Weighted gap variance within sector-month & 94.8\% \\
\addlinespace
\multicolumn{2}{l}{\textit{Panel B: raw-source and update validation}} \\
Raw-to-model mismatches, IAS / Weighted & 0 / 0 \\
Paired observations with same snapshot, row, and entity & 29,572 / 29,572 \\
IAS adjacent pairs unchanged & 94.05\% \\
Weighted adjacent pairs unchanged & 72.61\% \\
Median unchanged run, IAS / Weighted & 12 / 3 months \\
Weighted changes while IAS is unchanged & 22.46\% \\
Both scores change in the same adjacent pair & 4.93\% \\
\bottomrule
\end{tabular}
\begin{minipage}{0.94\linewidth}
\footnotesize
\textit{Notes:} Panel A compares IAS and WAKIS where both are observed; GICS sector is the finest historical peer-group field available in the analysis data. Panel B validates both accepted fields against the raw monthly MSCI source. Each field reproduces without unexplained mismatch, and paired observations use the same prior-month source row and mapped entity/security. The cadence difference is consistent with MSCI's published hierarchy and historical update guidance \citep{msci2024methodology,msci2022issuerfaq}. The diagnostics do not identify a single provider mechanism or rank the scores.
\end{minipage}
\end{table}

\FloatBarrier
\subsection{Observed-change timing synchronization diagnostic}

Table~\ref{tab:timing_sync} reports the bounded timing diagnostic motivated by the documented difference in score updating. A provider-native \texttt{iva\_rating\_date} field is present in the available historical MSCI extract, but it maps to only 22,238 of the 29,572 frozen paired observations (75.2\%) and the extract ends in February 2024, with no 2025 coverage. Using it would therefore change the sample period. To preserve the exact paired sample and specification, the diagnostic instead constructs a Weighted Average proxy that refreshes only when the observed IAS value changes, with contemporaneous initialization after missing-score resets and no future information. The proxy is explicitly an observed-change synchronization device, not a reconstruction of all MSCI rating actions. No alternative timing rule is searched.

\begin{table}[H]
\centering
\caption{Observed-change timing synchronization diagnostic for the paired MSCI score comparison.}
\label{tab:timing_sync}
\begin{minipage}{0.98\linewidth}
\centering
\scriptsize
\begin{tabular}{lrrrrr}
\toprule
Contrast / diagnostic & Baseline & Synchronized & Wild $p$ & 3-state Holm & 4-test Holm \\
\midrule
\multicolumn{6}{l}{\textit{Panel A: paired standardized contrasts}} \\
$\Delta_D$ & -0.2536 & -0.2859 & 0.0273 & -- & 0.0640 \\
$\Delta_0$ & 0.00263 & 0.00472 & 0.1847 & 0.1847 & 0.1847 \\
$\Delta_{-5}$ & -0.01005 & -0.00958 & 0.0160 & 0.0480 & 0.0640 \\
$\Delta_{-10}$ & -0.02273 & -0.02387 & 0.0189 & 0.0480 & 0.0640 \\
\addlinespace
\multicolumn{6}{l}{\textit{Panel B: synchronization diagnostics}} \\
Raw Weighted updates suppressed & \multicolumn{5}{r}{83.50\%} \\
Synchronized Weighted adjacent-pair change rate & \multicolumn{5}{r}{5.33\%} \\
Median unchanged run, synchronized Weighted & \multicolumn{5}{r}{12 months} \\
Attenuation of $|\Delta_{-5}|$ & \multicolumn{5}{r}{4.73\%} \\
\bottomrule
\end{tabular}
\par\vspace{2pt}
\raggedright\footnotesize
\textit{Notes:} The synchronized analysis preserves the exact paired sample, standardization rule, controls, fixed effects, and inference design. WAKIS refreshes only when the observed IAS score changes, with resets after missing-score observations and no future information. This lowers its adjacent-pair change rate from 27.39\% to 5.33\%; IAS changes in 5.95\% of pairs. The $-5\%$ contrast attenuates by 4.73\%. Wild inference and Holm families follow Table~\ref{tab:architecture_comparison}. Because provider-native rating dates do not span the full sample, this proxy tests observed cadence rather than reconstructing all MSCI rating actions or identifying a specific provider mechanism.
\end{minipage}
\end{table}

\FloatBarrier
\subsection{Final architecture validation}

Table~\ref{tab:architecture_validation} reports the bounded sector-time and leave-one-month-out validations of the paired architecture result. Panel A replaces ordinary month fixed effects with GICS sector-by-month fixed effects on the exact 29,572-observation paired sample while preserving firm fixed effects and firm-by-calendar-month clustering. Panel B returns to the accepted ordinary-month specification and omits each calendar month once while holding the full-sample score standardization fixed. The validation is evaluated as effect preservation and influence sensitivity rather than as a new search for favorable specifications.

\begin{table}[H]
\centering
\caption{Final sector-time and leave-one-month-out validation of the paired ESG-architecture comparison.}
\label{tab:architecture_validation}
\scriptsize
\begin{tabular}{llrrrrr}
\toprule
Panel & Quantity & Estimate & Wild $p$ & 3-state Holm & 4-test Holm & Baseline attenuation \\
\midrule
\multicolumn{7}{l}{\textit{Panel A: firm FE plus GICS sector-by-month FE}} \\
A & $\Delta_D$ & -0.1539 & 0.021 & -- & 0.049 & 39.3\% \\
A & $\Delta_0$ (pp) & 0.026 & 0.910 & 0.910 & 0.910 & -- \\
A & $\Delta_{-5}$ (pp) & -0.743 & 0.013 & 0.037 & 0.049 & 26.0\% \\
A & $\Delta_{-10}$ (pp) & -1.513 & 0.012 & 0.037 & 0.049 & 33.4\% \\
\addlinespace
\multicolumn{7}{l}{\textit{Panel B: ordinary-month FE leave-one-month-out influence}} \\
B & $\Delta_D$ & full: -0.2536 & -- & -- & -- & max: 26.6\% \\
B & $\Delta_D$ LOO range & $[-0.2807,-0.1862]$ & -- & -- & -- & sign: 83/83 \\
B & $\Delta_{-5}$ (pp) & full: -1.005 & -- & -- & -- & max: 22.3\% \\
B & $\Delta_{-5}$ LOO range (pp) & $[-1.180,-0.781]$ & -- & -- & -- & sign: 83/83 \\
\bottomrule
\end{tabular}
\begin{minipage}{0.97\textwidth}
\footnotesize
\textit{Notes:} Panel A uses the exact paired sample and full-sample standardization from Table~\ref{tab:architecture_comparison}, replacing month fixed effects with GICS sector-by-month fixed effects while retaining firm fixed effects and the same paired wild-cluster protocol. Panel B returns to the baseline month specification and omits each of the 83 months once without re-standardizing ESG. All omissions preserve the negative signs of $\Delta_D$ and $\Delta_{-5}$; the largest attenuation of $|\Delta_{-5}|$ is 22.3\%. The exercise evaluates coefficient preservation and influence, not a new significance search.
\end{minipage}
\end{table}

\FloatBarrier
\subsection{Post-primary period sensitivity}

Table~\ref{tab:covid_sensitivity} reports the single frozen 2020--2021 exclusion sensitivity undertaken after the architecture analysis. The original Joint80 thresholds and full-sample architecture standardization scales are retained. Baseline IAS and architecture differences remain directionally recognizable but lose finite-cluster precision, and no alternative exclusion window is searched.

\begin{table}[H]
\centering
\caption{Post-primary 2020--2021 exclusion sensitivity.}
\label{tab:covid_sensitivity}
\small
\begin{tabular}{llrrrr}
\toprule
Quantity & Sample & Estimate & Wild $p$ & 3-state Holm & 4-test Holm \\
\midrule
IAS ESG $\times$ Down & Full & -0.0702 & 0.157 & -- & -- \\
IAS ESG $\times$ Down & Exclude 2020--2021 & -0.0549 & 0.290 & -- & -- \\
\addlinespace
Architecture $\Delta_D$ (1 SD) & Full & -0.2536 & 0.028 & -- & 0.066 \\
Architecture $\Delta_D$ (1 SD) & Exclude 2020--2021 & -0.1972 & 0.097 & -- & 0.388 \\
Architecture $\Delta_{-5}$ (pp) & Full & -1.005 & 0.016 & 0.049 & 0.066 \\
Architecture $\Delta_{-5}$ (pp) & Exclude 2020--2021 & -0.725 & 0.155 & 0.309 & 0.388 \\
Architecture $\Delta_{-10}$ (pp) & Full & -2.273 & 0.018 & 0.049 & 0.066 \\
Architecture $\Delta_{-10}$ (pp) & Exclude 2020--2021 & -1.711 & 0.103 & 0.309 & 0.388 \\
\bottomrule
\end{tabular}
\begin{minipage}{0.97\textwidth}
\footnotesize
\textit{Notes:} Outcome months from 2020--2021 are removed while the original Joint80 thresholds and full-sample architecture standardization are retained. The reduced IAS and paired samples contain 21,520 and 21,366 observations, respectively, across 59 months. The 3-state and 4-test families follow Table~\ref{tab:architecture_comparison}. Seven remaining months reach $-5\%$ or below and none reaches $-10\%$; the latter contrast is therefore outside realized support.
\end{minipage}
\end{table}

\FloatBarrier
\bibliographystyle{plainnat}
\bibliography{references}

\end{document}